\documentclass[aps,twocolumn,pra,tightenlines,floatfix,showpacs,superscriptaddress]{revtex4-1}
\usepackage{graphicx}
\usepackage{amsmath}
\usepackage{amssymb}
\usepackage{times}
\usepackage[english]{babel}
\newcommand{\Tr}[1]{\mathrm{Tr} [ \, #1 \, ]} 
\begin{document}

\title{Quench Dynamics and Emergence of Phase Separation in Two-Component Atomic Bose Gases at Zero Temperature and above the BEC Critical  Temperature}

\author{Chih-Chun Chien}
\affiliation{Theoretical division, Los Alamos National Laboratory, Los Alamos, NM 87545, USA.}
\author{Fred Cooper}
\affiliation{Theoretical division, Los Alamos National Laboratory, Los Alamos, NM 87545, USA.}
\affiliation{Santa Fe Institute, Santa Fe, NM 87501, USA.}

\begin{abstract}
We study the dynamics of two-component atomic Bose gases initially in a mixture encountering a sudden quench of the inter-species interactions. The  dynamics above the critical temperature $T_c$ is studied using a leading order large-N approximation that predicted a phase transition from mixing to phase separation as a function of the inter-species coupling. Here we explore the dynamics of this phase transition following a quench and compare our results to those found at zero temperature using the time-dependent Gross-Pitaevskii equations which ignore quantum and thermal fluctuations. In the regime above $T_c$ where no condensate is present, however, the  time evolution of the densities following the quench exhibits features similar to that found at zero temperature where only the condensates contribute to the densities. When the inter-species interaction jumps above the critical value, we observe dynamical transitions from a homogeneous mixture to a phase-separated structure for both cases. Our simulations suggest that at temperatures above $T_c$ where no condensate is present this dynamical transition should still be observable in experiments.
\end{abstract}

\pacs{05.30.Jp,67.85.-d,03.75.Kk,03.75.Mn}

\maketitle

Experiments on tuning the interactions \cite{2BECexp1,2BECexp2} and monitoring the dynamics \cite{2BECexp_dyna1,2BECexp_dyna2} of two-component atomic Bose-Einstein condensates (BECs) have led to observations of interesting phenomena and inspired studies using such highly controllable systems. The dynamics of two-component Bose gases may simulate interesting phenomena in cosmology \cite{Fischer04,Visser06}. In the weak-coupling regime at zero temperature ($T=0$), two-component Bose gases have been broadly studied using mean-field Bogoliubov theory showing a mixture to phase-separation transition when the inter-species interaction exceeds a critical value determined by the intra-species interactions \cite{FetterJLTP78,EddyPRL98,PethickBEC}. 

The dynamics of multi-component BECs have been simulated using the time-dependent Gross-Pitaevskii (TDGP) equations \cite{SinatraPRL99,Kasamatsu04,*Kasamatsu06,ZurekPRL11}, which are restricted to $T=0$ and weak coupling due to the approximations involved. The stability prediction of the Bogoliubov theory (which is a one-loop correction to the condensate contribution to the partition function) is the same as that found by linear stability analysis of the TDGP equations. Self-consistent mean field theories allow one to study the dynamics at finite temperature as well as in stronger coupling regimes. Previously we have studied both the large-$N$ expansion \cite{OurlargeN} and the leading-order-auxiliary-field (LOAF) theory \cite{LOAFPRL} for single-component Bose gases \cite{LOAF_Tcnote}. The LOAF theory becomes identical to the leading-order large-N expansion when there is no condensate. Thus the latter is appropriate for studying  the dynamics of the phase-separation transition in the regime above the critical temperature $T_c$, which has been less explored in the literature.

A leading-order large-N approximation of two-component Bose gases shows that there is a mixture to phase-separation transition in equilibrium \cite{OurlargeN}. Above $T_c$, there is no condensate and the transition is driven by a competition among the kinetic energy, intra-, and inter- species interactions. The goal of this paper is to study the dynamics of the system across this structural phase transition above $T_c$ and compare it to the results of a simulation at $T=0$ using the TDGP equations. We consider the case where an initial mixture of the two species  evolves into a phase-separated structure after a sudden quench in the inter-species interactions.

We  first discuss  zero temperature simulations of the quench using the TDGP equations  to review what is know in that case, to benchmark our numerics, as well as to be able to compare with the finite temperature calculation. We will see below that the mode functions determining the correlation functions of the large-N approximation obey similar partial differential equations to those found for the condensates in the TDGP equations. 
We begin with the one-dimensional coupled TDGP equations \cite{PethickBEC} modeling a condensate at $T=0$ in a torus 
\begin{eqnarray}\label{eq:2comp_TDGP}
i\hbar\frac{\partial \phi_j}{\partial t}&=&-\frac{\hbar^2}{2m_j}\frac{\partial ^2\phi_j}{\partial x^2}+\lambda_{jj}|\phi_j|^2\phi_j+\lambda_{j\bar{j}}|\phi_{\bar{j}}|^2\phi_j.
\end{eqnarray}
Here $\phi_j$, $j=1,2$, denotes the condensate field for species $j$, $\bar{j}=1$ if $j=2$ and $\bar{j}=2$ if $j=1$, and the direction $x$ is along the torus. We consider periodic boundary conditions and further assume $\rho_1=\rho_2=\rho_0$ and $m_1=m_2=m$. The unit of time is $t_0=2m/(\hbar\rho_0^{2/3})$ and the unit of length is $\rho_0^{-1/3}\equiv k_0^{-1}$. The coupling constants are related to the $s$-wave scattering lengths by $\lambda_{ij}=4\pi\hbar^2 a_{ij}/m$ for $i,j=1,2$. We will set $\hbar\equiv 1$. 
The TDGP equations can be solved using the split-step Fourier method \cite{Taha,Bao}. We calibrate the grid size in real space and the time increment in our simulations using the exact solutions of Ref.~\cite{FredNLSE02}.

\begin{figure}
\begin{centering}
\includegraphics[width=2.8in]{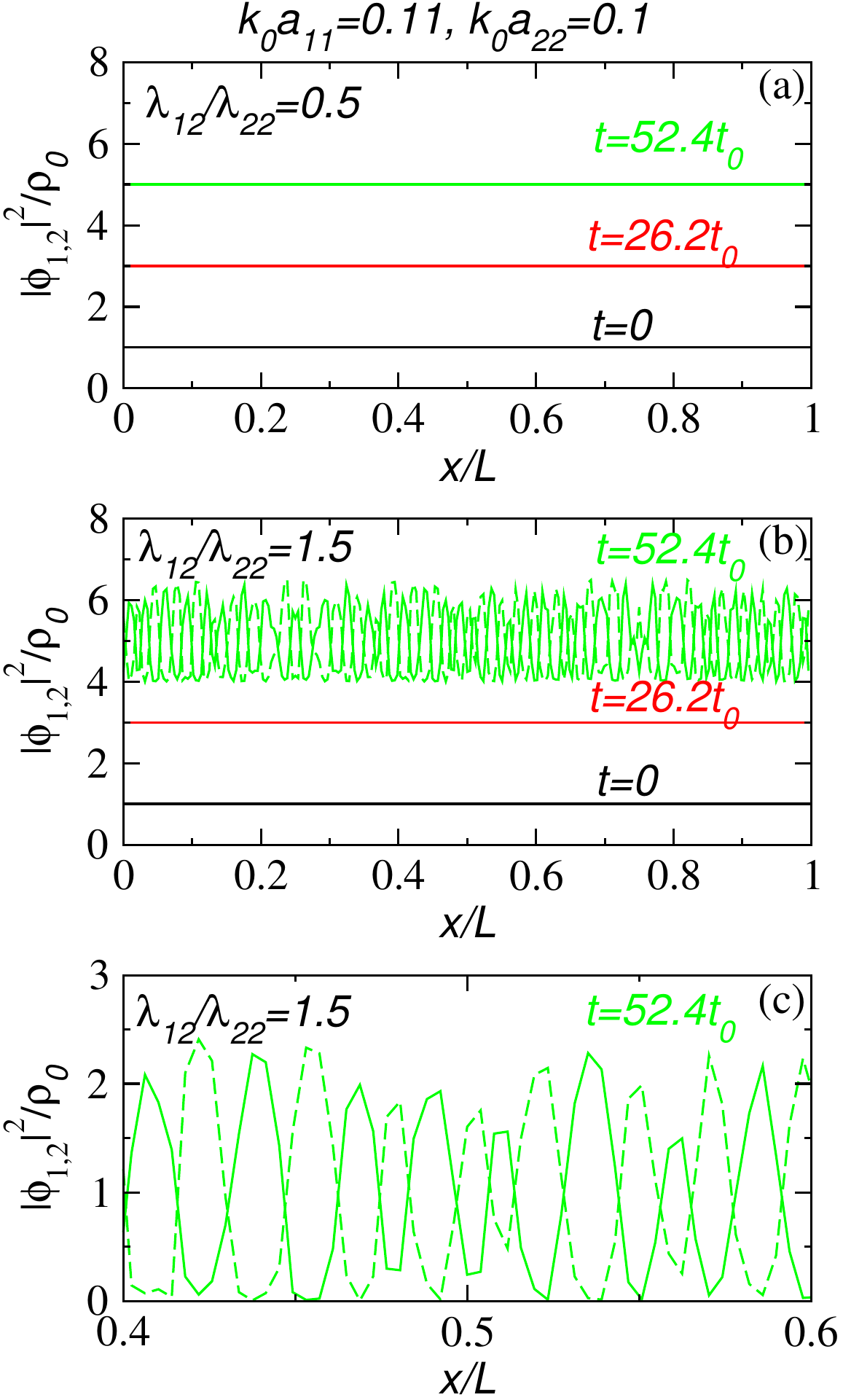}
\par\end{centering}
\caption{(Color online) Evolution of the density profiles for two-component BEC according to the TDGP equations \eqref{eq:2comp_TDGP}. The initial state is at $T=0$ and $k_0 a_{11}=0.11$, $k_0 a_{22}=0.1$. The profiles are offset by $2$ for each time increment and we show $t=0, 26.2t_0, 52.4t_0$. (a) $\lambda_{12}/\lambda_{22}=0.5$. (b) $\lambda_{12}/\lambda_{22}=1.5$. The structure of (b) is shown in more detail in (c) for $t=52.4t_0$ without offset. The solid (dashed) line correspond to species 1 (2).}
\label{fig:2BECT0}
\end{figure}

For two-component BECs in equilibrium at $T=0$, a structural phase transition occurs when $\lambda_{12}^2/(\lambda_{11}\lambda_{22}) >1$ \cite{EddyPRL98,PethickBEC}. We consider the case where $\lambda_{12}^2/(\lambda_{11}\lambda_{22}) <1$ initially so the system is a mixture. Then the inter-species interaction suddenly changes to $\lambda_{12}^2/(\lambda_{11}\lambda_{22}) >1$. Figure~\ref{fig:2BECT0} shows the dynamics of the density profiles $|\phi_1|^2$ and $|\phi_2|^2$ for (a) the case without any change of $\lambda_{12}$ and for (b) the case with a sudden quench in $\lambda_{12}$. We assume $\lambda_{11}$ and $\lambda_{22}$ are fixed and choose their values to be close to each other, which is an approximation for the systems discussed in Refs.~\cite{2BECexp1,2BECexp2}. In the initial density profiles we impose a small perturbation \cite{T0note}. Without a quench in $\lambda_{12}$ the profiles evolve smoothly with no observable structure. When $\lambda_{12}$ suddenly  jumps above the critical value, the two components start to form domains and exclude each other as shown in Fig.~\ref{fig:2BECT0} (b) and (c). The large number of domains emerging during the dynamics suggests that the initial mixture state is a highly excited state when compared to the genuine ground state of the large value of $\lambda_{12}$. Ref.~\cite{Kasamatsu04,*Kasamatsu06} investigated more dynamics using TDGP equations.

In the BEC phase, to go beyond the zero-temperature and weak-coupling limit of the TDGP equations, one should include the pair correlations discussed in Refs.~\cite{LOAFPRL,LOAFlong} as well as the fluctuations of the normal density included in the large-N approximation. The interplay among the condensate, quasi-particles, and pair correlations during the dynamics will be an interesting topic for future studies.

In the finite-temperature quantum regime above $T_c$, all condensates vanish and only the fluctuations of the normal density are important.  The relevant mean-field theory which preserves symmetries is the leading term in the large-N expansion of the theory, which for  the normal phase of a two-component Bose gas was presented in Ref.~\cite{OurlargeN}. There we focused on the static properties of the theory and the partition function which was  relevant to study the phase diagram. 
 Here we study the dynamics coming from the effective action derived in Ref.~\cite{OurlargeN}.
The action describing two-component bosons with contact interactions is 
\begin{eqnarray}
S&=&\int [dx]\left\{\sum_{j=1,2}\left[ \phi^{*}_{j}i\partial_{t}\phi_{j} + \phi^{*}_{j}(\frac{\nabla^{2}}{2m_j})\phi_{j}-\mu_{j}\phi^{*}_{j}\phi_{j}- \right.\right.\nonumber \\
& &\left.\left. \frac{1}{2}\lambda_{jj}(\phi^{*}_{j}\phi_{j})^{2}\right]-\lambda_{12}(\phi^{*}_{1}\phi_{1})(\phi^{*}_{2}\phi_{2})\right\}.
\end{eqnarray} 
Here $[dx]\equiv dt d^{d}x$. We introduce the large parameter $N$ into the theory by the replication trick $\phi_j \rightarrow \phi_{j,n}$ where $n =1, 2, \ldots N$ and rescale the coupling constants $\lambda_{jj} \rightarrow \lambda_{jj}/N$ and $\lambda_{12}\rightarrow \lambda_{12}/N$ \cite{OurlargeN}. 
The generating functional for the correlation functions is given by 
$Z[J] = \int \left(\prod_{j=1,2; n=1}^{N}\mathcal{D}\phi_{j,n}\mathcal{D}\phi_{j,n}^*\right)  e^{ iS[J,   \phi_{j,n},  \phi_{j,n}^*]}$.
Here the action with the source term after the replication becomes
\begin{eqnarray}
S&=&-\int [dx] \left[\frac{1}{2}\Phi^{\dagger}\tilde{G}^{-1}_{0}\Phi+\sum_{j=1,2}\frac{\lambda_{jj}}{2N}\left(\sum_{n=1}^{N}\phi_{j,n}^*\phi_{j,n}\right)^2+ \right. \nonumber \\
& & \frac{\lambda_{12}}{N}\left(\sum_{n=1}^{N}\phi_{1,n}^* \phi_{1,n}\right)\left. \left(\sum_{n=1}^{N}\phi_{2,n}^* \phi_{2,n}\right)-J^{\dagger}\Phi \right].
\end{eqnarray}
Here $\Phi=(\phi_{1,1},\phi_{1,1}^{*},\phi_{2,1},\phi_{2,1}^{*}, \cdots)^{T}$, $\bar{G}_{0}^{-1}=diag(h_1^{(+)},h_1^{(-)},h_2^{(+)},h_2^{(-)},\cdots)$, $\tilde{G}^{-1}_{0}=\bar{G}_{0}^{-1}- diag(\mu_1,\mu_1,\mu_2,\mu_2,\cdots)$ is the bare Green's function, and $h_{j}^{(\pm)}=\mp i\partial_{t}-\nabla^{2}/(2m_j)$ for $j=1,2$. 
There are $N$ copies in $\Phi$, $\bar{G}_{0}^{-1}$, and $\tilde{G}_{0}^{-1}$. $J$ is the source coupled to $\Phi$.

We next insert the identity 
\begin{equation}
1=\int\left(\prod_{j=1}^{2}\mathcal{D}\chi_j\mathcal{D}\alpha_j\right) e^{ \int [dx]\sum\limits_{j=1}^{2}\left[\frac{N\chi_j}{\lambda_{jj}}(\alpha_j-\frac{\lambda_{jj}}{N}\sum\limits_{n=1}^{N}\phi_{j,n}^* \phi_{j,n})\right]  } \nonumber
\end{equation} inside the path integral for Z[J] which
introduces two functional delta functions \cite{MosheLargeN} so one can replace $\sum_{n=1}^{N}\phi_{j,n}^* \phi_{j,n}$ by $(N/\lambda_{jj})\alpha_j$ in $S$. Here the $\chi$ integration contour runs parallel to the imaginary axis (see Ref.~\cite{MosheLargeN}). This replacement facilitates our resummation scheme and we will treat $1/N$ as a small parameter. Let $G_{0}^{-1}\equiv\bar{G}_{0}^{-1}+diag(\chi_1,\chi_1,\chi_2,\chi_2,\cdots)$. After performing the Gaussian integral in the $\phi_{j,n}$, one has $S_{eff}=$
\begin{eqnarray}
& &\int [dx] \left[\frac{J^{\dagger}G_{0}J}{2}+\sum_{j=1}^{2}\left(\frac{N}{\lambda_{jj}}\mu_j\alpha_j-\frac{N\alpha_j^2}{2\lambda_{jj}}+\frac{N}{\lambda_{jj}}\chi_j\alpha_j \right. \right)-\nonumber \\
& & \frac{N\lambda_{12}}{\lambda_{11}\lambda_{22}}\alpha_1\alpha_2 -\left. \frac{1}{2}Tr\ln G_{0}^{-1} +K^{\dagger}X\right].
\end{eqnarray}
Here $X=(\chi_1,\chi_2,\alpha_1,\alpha_2)^{T}$ with its source term $K$  and expectation value $X_c$.
We evaluate the path integrals over $\chi_j, \alpha_j$ via the method of stationary phase 
and at leading order we keep only the contributions at the stationary phase point.  

The generating functional of the one-particle irreducible diagrams is obtained from $\ln Z$ by Legendre transform: $\Gamma=\int [dx] (J^{\dagger}\Phi_c+ K^{\dagger} X_c+i\ln Z)$, where $\Phi_c$ and $X_c$ are the classical values of $\Phi$ and $X$ respectively.  Keeping the leading term in the $1/N$ expansion and then setting $N=1$, we obtain  
\begin{eqnarray}
\Gamma&=&\int [dx]\left\{\frac{\Phi^{\dagger}G_0^{-1}\Phi}{2}-\sum_{j=1}^{2}\left(\frac{\mu_j\alpha_j}{\lambda_{jj}}-\frac{\alpha_j^2}{2\lambda_{jj}}+ \frac{\chi_j\alpha_j}{\lambda_{jj}} \right)+\right.\nonumber \\
& & \frac{\lambda_{12}}{\lambda_{11}\lambda_{22}}\alpha_1\alpha_2  + \left. \frac{1}{2}Tr\ln G_{0}^{-1} \right\}.
\end{eqnarray}
Here $\Phi=(\phi_1,\phi_1^{*},\phi_2,\phi_2^{*})^{T}$ and $G_{0}^{-1}$ has been reduced to a $4\times 4$ matrix. The equation for the condensate in the presence of sources is given by  $\delta \Gamma/\delta \Phi^\dag  = J=\int [dx] G_{0}^{-1}\Phi$.

For the initial conditions of our simulations we consider static homogeneous fields and define the effective potential as $V_{eff}=\Gamma/(NV \bar{t})$, where $\bar{t}$ is the length of time in the integral.
The broken-symmetry condition is determined from the true minimum of the effective potential:  $\delta V_{eff}/\delta \phi_j^*=0$, which becomes $\chi_j\phi_j=0$. In the normal phase $\phi_j=0$ while in the broken-symmetry phase $\chi_j=0$.  In the normal phase, the term $\frac{1}{2}\Phi^{\dagger}G_0^{-1}\Phi$ is zero at the minimum of the potential (which occurs at $\phi_j =0$). 
The condition $\delta V_{eff}/\delta \alpha_j=0$ fixes the relation between $\alpha_j$ and $\mu_j$. Thus 
\begin{eqnarray}\label{eq:chi_eq}
\chi_j=-\mu_j+\alpha_j+\frac{\lambda_{12}}{\lambda_{\bar{j}\bar{j}}}\alpha_{\bar{j}}; ~~ \alpha_j=\lambda_{jj} \langle\phi_j^*\phi_j \rangle .
\end{eqnarray}
Here the expectation value means 
$ \Tr{ \rho_0 \phi_j^*\phi_j }$, where $\rho_0$ is the density matrix describing the quantum state at the initial time. The parameters are determined self-consistently at the initial time (see below). In the BEC phase the expectation value would include the condensates plus the quantum fluctuations.

To describe the dynamics, we derive the equations of motion from the identity $\int dz G^{-1}(\textrm{x},\textrm{z})G(\textrm{z},\textrm{y})=\delta_{C}(\textrm{x}-\textrm{y})$, where $G^{-1}=\delta^2 \Gamma/\delta\phi_{a}\delta\phi_{b}$ and we use the closed time path (CTP) boundary conditions on the Green's functions due to  Schwinger and Keldysh (S-K)  (see \cite{FredPRD03,Kamenevbook} and references therein). Here $\phi_{a,b}=\phi_{j}, \phi_{j}^{*}$, $\delta_C(\textrm{x}-\textrm{y})$ is a four-dimensional delta function defined on the S-K contour, and $\textrm{x},\textrm{y},\textrm{z}$ denote four-vectors. One finds that $G^{-1}_{jj}(\textrm{x},\textrm{y})=\delta_C(\textrm{x}-\textrm{y})(h^{(+)}+\chi_j)$ and the off-diagonal elements vanish. Following Ref.~\cite{FredPRD03}, the equations for the Green's functions can be satisfied by introducing a set of fields $\phi_j(\textrm{x})$ satisfying the homogeneous equation $\int d\textrm{y} G_{jj}^{-1}(\textrm{x},\textrm{y})\phi_j(\textrm{y})=0$, where $d\textrm{y}\equiv dtd^{d}y$. Explicitly, 
\begin{equation}\label{eq:phiEOM}
i\partial_t\phi_j=(-\partial_{x}^{2}-\mu_j+\alpha_j+\frac{\lambda_{12}}{\lambda_{\bar{j}\bar{j}}}\alpha_{\bar{j}}.) \phi_j.
\end{equation}
Here  $\alpha_j=\lambda_{jj} \langle\phi_j^*\phi_j \rangle$ and the expectation value is over the initial (thermal) state of the system.
This is to be compared with Eq. (\ref{eq:2comp_TDGP}) for the $T=0$ case. 

\begin{figure}
\begin{centering}
\includegraphics[width=2.8in]{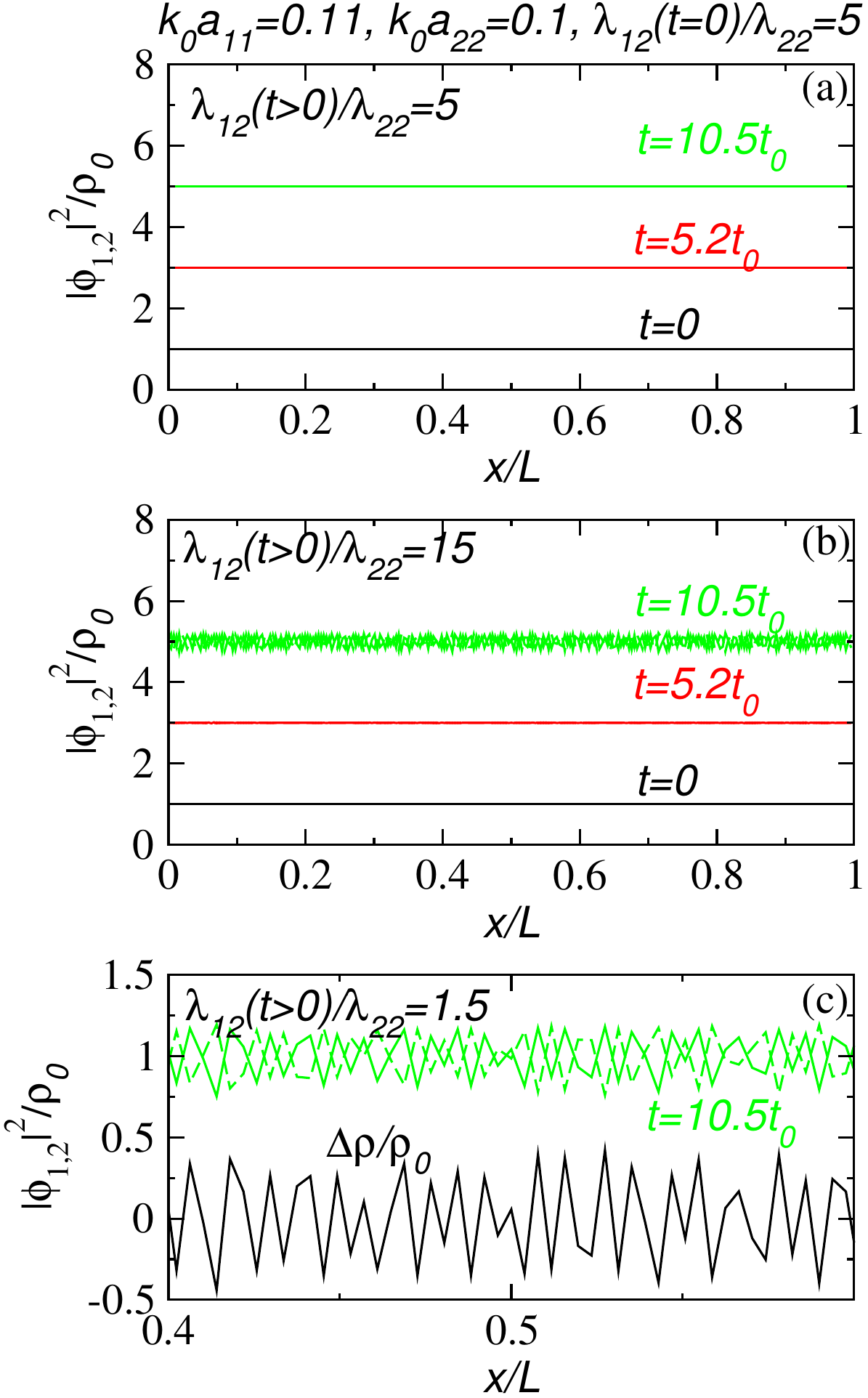}
\par\end{centering}
\caption{(Color online) Evolution of the density profiles for species 1 (solid lines) and 2 (dashed lines). Here $k_0 a_{11}=0.1$, $k_0 a_{22}=0.11$, $T=3T_0$, and $N=256$. Initially $k_0 a_{12}(t=0)=0.5$. The profiles for $t>0$ is offset by $2$ after each time increment. We show the profiles at $t=0$, $5.2t_0$, and $10.5t_0$. (a) shows $k_0 a_{12}(t>0)=0.5$ below the critical value. (b) shows $k_0 a_{12}(t>0)=1.5$ above the critical value. The detailed structure of (b) is shown in (c) and the black line shows the density difference $\Delta \rho=\rho_1-\rho_2$.}
\label{fig:2N256}
\end{figure}

To implement the initial condition on the Green's function,  one represents the quantum field $\phi_i$  and its conjugate by a mode expansion in terms of canonical annihilation and creation operators: 
$
\phi_j(\textrm{x})=\sum_{k=1}^{N} a_{jk}f_{jk}(x,t).
$
The operator $a_{jk}$ is the time independent  initial value of the annihilation operator in the Heisenberg picture. The temperature enters by specifying that the initial distribution for each species is given by a Bose-Einstein distribution at a given temperature.  Thus  $\langle a^{\dagger}_{jk}a_{j^{\prime}k^{\prime}}\rangle=\delta_{jj^{\prime}}\delta_{kk^{\prime}}[\exp(\omega_{j0}/k_B T)-1]^{-1}$, where $\omega_{j0}=k^2/(2m_j)+\chi_j(t=0)$.
In our simulations we focus on the motion along the direction along the torus and neglect the details in the other two spatial dimensions. 
The time-ordered Green's function of the $j$-th species is then written in terms of these fields as
$G_{jj}(\textrm{x},\textrm{y})=-i\langle \mathcal{T}_C [\phi_j(\textrm{x})\phi_j^*(\textrm{y})]\rangle$, or explicitly $\theta_C(\textrm{x}_0-\textrm{y}_0)G_{>,j}(\textrm{x},\textrm{y})+\theta_C(\textrm{y}_0-\textrm{x}_0)G_{<,j}(\textrm{x},\textrm{y})$.
Here $\mathcal{T}_C$ is the CTP time-ordered product and $\theta_C(\textrm{x}_0)$ is the CTP step function while $G_{>,j}(\textrm{x},\textrm{y})=-i\langle \phi_j(\textrm{x})\phi_j^{*}(\textrm{y})\rangle$ and $G_{<,j}=i\langle \phi_j^*(\textrm{x})\phi_j(\textrm{y})\rangle$ following the convention of Ref.~\cite{Kamenevbook}. In terms of the fields $\phi_j(\textrm{x})$, we have  $\rho_j(\textrm{x})=\langle \phi_j^*(\textrm{x})\phi_j(\textrm{x})\rangle$. 

In our simulations, we follow $N$ modes, where $N$ is also the number of grid points in $x$. The initial conditions are chose as $f_{jk}(x,0)=\sqrt{1/N_{j0}}\exp[i(kx-\omega_{j0}t)]$, where the normalization coefficients are chosen to satisfy $\rho_j(x,t=0)=|\phi_{j}(x,t=0)|^{2}=\rho_0=k_0^{3}$ for $j=1,2$. The equations of motion \eqref{eq:phiEOM} then lead to
$i\partial_t f_{jk}(x,t)=[-\partial_{x}^2+\chi_j(x,t)]f_{jk}(x,t)$.
The densities are evaluated by 
$\rho_j(x,t)=\langle \phi_j^*(x,t)\phi_j(x,t)\rangle=\sum_{k}f_{jk}^*(x,t)f_{jk}(x,t)n_{jk}(T)$. 
The composite fields are then evaluated by $\alpha_j(x,t)=\lambda_{jj}\rho_j(x,t)$ and
$\chi_j(x,t)=-\mu_j+\alpha_j(x,t)+\frac{\lambda_{12}}{\lambda_{\bar{j}\bar{j}}}\alpha_{\bar{j}}(x,t)$.
Here we assume the relation between $\alpha_j(x,t)$ and $\chi_j(x,t)$ follows the equilibrium relation \eqref{eq:chi_eq} but the values of $\mu_j$ are fixed by the equilibrium values so that the evolution is continuous.

We again consider the setup where initially the system is a mixture described by a thermal distribution at temperature $T> T_c$ and suddenly $\lambda_{12}$ jumps above the critical value. The initial condition is determined by solving the equilibrium equations of state $\rho_j=-\partial V_{eff}/\partial \mu_{j}$ and Eq.~\eqref{eq:chi_eq} \cite{OurlargeN}: 
\begin{eqnarray}\label{eq:EOS}
\rho_j=\sum_{p}n(\omega_j), ~~\alpha_j(t=0)=\lambda_{jj}\rho_j.
\end{eqnarray}
Here $n(x)=[\exp(x/k_B T)-1]^{-1}$ , $\omega_j=p^{2}/(2m_j)+\chi_{j}(t=0)$, and $p$ is a continuous variable representing the momentum. We choose $\rho_1=\rho_2=\rho_0$ and $m_1=m_2=m$. Eq.~\eqref{eq:EOS} gives the values for $\chi_j$ and $\alpha_j$, and $\mu_j$ can be determined from Eq.~\eqref{eq:chi_eq}. For $k_0 a_{11}=0.11$, $k_0 a_{22}=0.1$, and $k_0 a_{12}(t=0)=0.5$, the system is stable in the mixture phase and the corresponding parameters are obtained. For illustrative purposes, the initial temperature is chosen as $T=3T_0$, where $T_0$ is the BEC temperature of an noninteracting Bose gas with density $\rho_0$ \cite{Tc_note}.  The dynamics when  $k_0 a_{12}(t>0)$ jumps to $1.5$ (above the critical value) is then simulated.

The evolution of the density profiles is shown in Figures~\ref{fig:2N256} (a) and (b) corresponding to the cases without and with a jump in $\lambda_{12}$ for $N=256$ modes. In contrast to the $T=0$ case, here we can use a uniform initial condition with $\rho_1=\rho_2$. One can see that if $\lambda_{12}$ remains below the critical value (which is close to $10\sqrt{\lambda_{11}\lambda_{22}}$), the system remains a mixture. In contrast, when $\lambda_{12}$ jumps above the critical value, the two density profiles evolve into domains and the domains of different species avoid each other. Therefore many interfaces separating different species emerge during the dynamics. We found that our results are robust when the number of modes double or when smaller time steps are taken. For larger jumps of $\lambda_{12}$ we found that the time it takes to form those domains decreases. If $T/T_0$ is too large, the kinetic energy dominates and one cannot find a structural transition for reasonable parameters. 
When compared to the results at $T=0$, one can see that the modulations of the densities at finite temperature above $T_0$  are not as strong. Therefore above $T_0$ we expect that density ripples corresponding to the phase-separated phase should emerge when $\lambda_{12}$ is above the critical value. The mixture background above $T_0$ is due to kinetic energy effects that are not significant at $T=0$. The density difference shown on Fig.~\ref{fig:2N256} should be large enough to be distinguished from the mixture background in experiments. For larger jumps of $\lambda_{12}$, the density ripples becomes more prominent due to larger inter-species interactions so experimentally one should use the largest available jump for better observations of this dynamical structural transition.

In summary, we found that a dynamical structural transition should be observable in two-component Bose gases at $T=0$ and temperatures above $T_c$ so the presence of BECs is not a crucial ingredient for this transition. Our formalism for finite $T$ dynamics provides a coherent description and complements conventional kinetic-theory approaches \cite{PK}. If optically controllable collisions (see references in Ref.~\cite{ChienOFR}) can be realized in two-component Bose gases, one may study more complicated structures and dynamics in the presence of inhomogeneous interactions.

The authors acknowledge the support of the U. S. DOE through the LANL/LDRD Program and the hospitality of Santa Fe Institute.

\bibliographystyle{apsrev4-1}

\begin{thebibliography}{28}
\bibitem [{LOA()}]{LOAF_Tcnote}%
  \BibitemOpen
  \href@noop {} {}\bibinfo {note} {The LOAF approach includes fluctuations of
  both the normal and anomalous densities and predicts a shift in the critical
  temperature $T_c$ from the noninteracting-gas value $T_0$. It is the same as
  that found in higher-order large-N expansion and in qualitative agreement
  with the Monte Carlo result \cite{LOAF_AL}.}\BibitemShut {Stop}%
\bibitem [{T0n()}]{T0note}%
  \BibitemOpen
  \href@noop {} {}\bibinfo {note} {The initial perturbation is of the form $\pm
  u\sin(2\pi x)$ in the two components with $u=10^{-8}\rho_0$ and we have
  checked that the evolution is not sensitive to the value of $u$ as long as it
  is small.}\BibitemShut {Stop}%
\bibitem [{Tc_()}]{Tc_note}%
  \BibitemOpen
  \href@noop {} {}\bibinfo {note} {Note that the BEC temperature is not shifted
  in the leading-order large-N approximation \cite{OurlargeN}.}\BibitemShut
  {Stop}%
\end{thebibliography}
%

\end{document}